\documentclass[8.5pt,twoside,twocolumn]{article}
\usepackage[super,sort&compress,comma]{natbib} 
\usepackage[version=3]{mhchem}
\usepackage{times}
\usepackage{sectsty}
\usepackage{balance} 
\usepackage{graphicx} 
\usepackage{lastpage}
\usepackage{float}
\usepackage[format=plain,singlelinecheck=false,font=small,labelfont=bf,labelsep=space]{caption} 
\usepackage{fancyhdr}
\usepackage{amssymb, amsmath}
\begin{document}
\thispagestyle{plain}
\fancypagestyle{plain}{
\renewcommand{\headrulewidth}{1pt}}
\renewcommand{\thefootnote}{\fnsymbol{footnote}}
\renewcommand\footnoterule{\vspace*{1pt}%
\hrule width 3.4in height 0.4pt \vspace*{5pt}} 
\setcounter{secnumdepth}{5}
\makeatletter 
\def\subsubsection{\@startsection{subsubsection}{3}{10pt}{-1.25ex plus -1ex minus -.1ex}{0ex plus 0ex}{\normalsize\bf}} 
\def\paragraph{\@startsection{paragraph}{4}{10pt}{-1.25ex plus -1ex minus -.1ex}{0ex plus 0ex}{\normalsize\textit}} 
\renewcommand\@biblabel[1]{#1}            
\renewcommand\@makefntext[1]%
{\noindent\makebox[0pt][r]{\@thefnmark\,}#1}
\makeatother 
\renewcommand{\figurename}{\small{Fig.}~}
\sectionfont{\large}
\subsectionfont{\normalsize} 
\fancyfoot{}
\fancyfoot[RO]{\footnotesize{\sffamily{\thepage}}}
\fancyfoot[LE]{\footnotesize{\sffamily{\thepage}}}
\fancyhead{}
\renewcommand{\headrulewidth}{1pt} 
\renewcommand{\footrulewidth}{1pt}
\setlength{\arrayrulewidth}{1pt}
\setlength{\columnsep}{6.5mm}
\setlength\bibsep{1pt}
\twocolumn[
  \begin{@twocolumnfalse}
\noindent\LARGE{\textbf{Co-nonsolvency of PNiPAM at the transition between solvation \\mechanisms}}
\vspace{0.6cm}

\noindent\large{\textbf{I. Bischofberger$^{\ast\ast}$, D. C. E. Calzolari and
V. Trappe$^{\ast}$}\vspace{0.5cm}}
\vspace{0.6cm}

\noindent \normalsize{We investigate the co-nonsolvency of poly-N-isopropyl acrylamide (PNiPAM) in different water/alcohol mixtures and show that this phenomenon is due to two distinct solvation contributions governing the phase behavior of PNiPAM in the water-rich and alcohol-rich regime respectively. While hydrophobic hydration is the predominant contribution governing the phase behavior of PNiPAM in the water-rich regime, the mixing contributions governing the phase behavior of classical polymer solutions determine the phase behavior of PNiPAM in the alcohol-rich regime. This is evidenced by distinct scaling relations denoting the energetic state of the aqueous medium as a key parameter for the phase behavior of PNiPAM in the water-rich regime, while the volume fractions of respectively water, alcohol and PNiPAM become relevant parameters in the alcohol-rich regime. Adding alcohol to water decreases the energetics of the aqueous medium, which gradually suppresses hydrophobic hydration, while adding water to alcohol decreases the solvent quality. Consequently, PNiPAM is insoluble in the intermediate range of solvent composition, where neither hydrophobic hydration nor the mixing contributions prevail. This accounts for the co-nonsolvency phenomenon observed for PNiPAM in water/alcohol mixtures.}
\vspace{0.5cm}
 \end{@twocolumnfalse}
  ]
\footnotetext{\textit{University of Fribourg, Department of Physics, CH-1700 Fribourg \\$^{\ast\ast}$current address: University of Chicago, Department of Physics, Chicago, IL 60637}} 
\footnotetext{\textit{E-mail:veronique.trappe@unifr.ch}}

\section{\label{sec:intro}Introduction}
Co-nonsolvency is a rather rare phenomenon, where a polymer perfectly soluble in two different solvents becomes insoluble in mixtures of both.\cite{Wolf1978} As a well-known example let us consider solutions of poly-N-isopropyl acrylamide \mbox{(PNiPAM)} in water/methanol mixtures.\cite{Schild1991, Winnik1990} At a fixed temperature of \mbox{$T$ = $20\,^{\circ}{\rm C}$} PNiPAM readily dissolves in both pure water and pure methanol, forming optically transparent solutions. However, mixing these solutions at certain proportions leads to the formation of precipitated phases, which is evidenced by the appearance of turbidity.\cite{Schild1991, Winnik1990, Costa2002} This impressive phenomenon is illustrated in the upper panel of Fig. 1, where we show a series of images taken for PNiPAM solutions in water/methanol mixtures with varying methanol molar fraction $X_\mathrm{MeOH}$. In the range of 0.13 $<$ $X_\mathrm{MeOH}$ $<$ 0.4 PNiPAM is insoluble, while for $X_\mathrm{MeOH}$ $<$ 0.13 and $X_\mathrm{MeOH}$ $>$ 0.4 PNiPAM is soluble.

\begin{figure}[htb]
	\centering
	\includegraphics[scale=0.65]{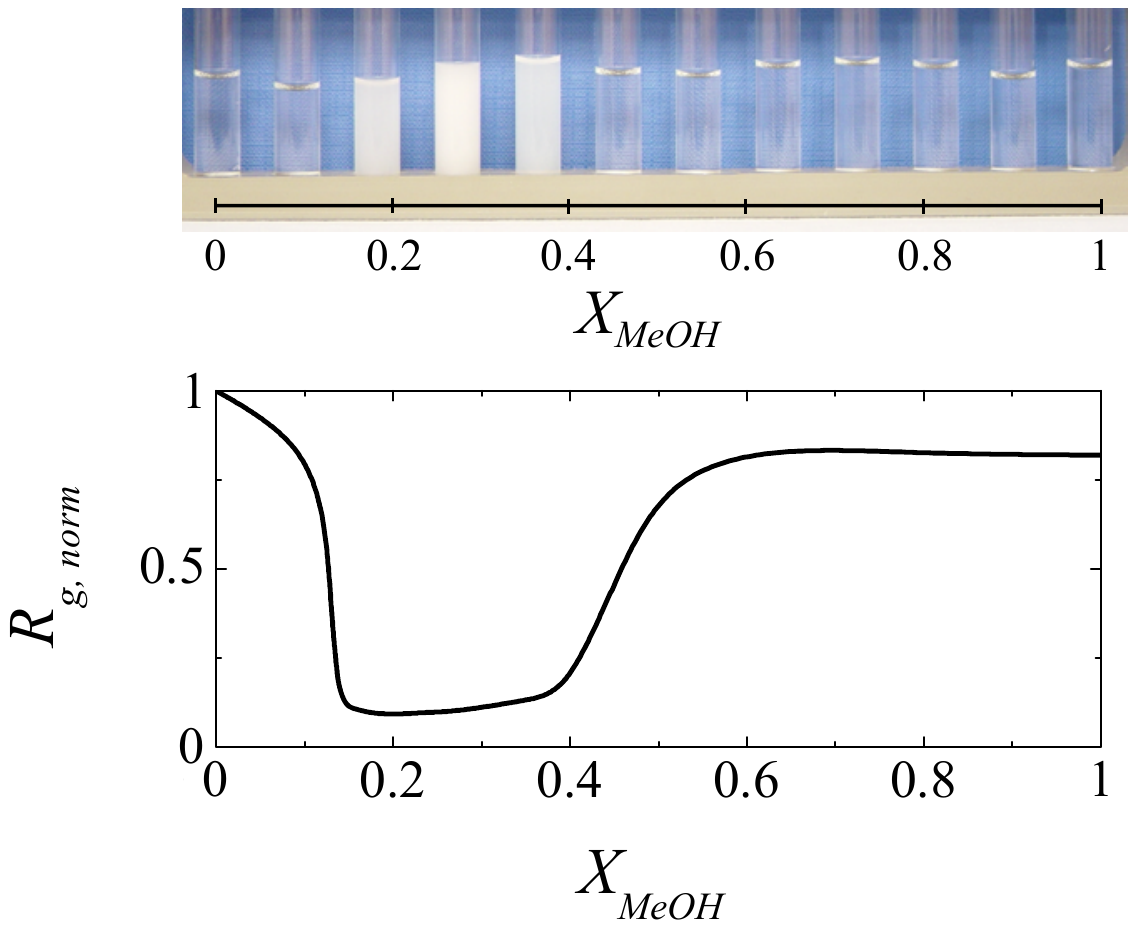}
	\caption{\label{Fig1}Co-nonsolvency effect and re-entrant coil-to-globule-to-coil transition of PNiPAM in water/methanol mixtures observed at a fixed temperature of $T$ = $20\,^{\circ}{\rm C}$. Upper panel: Phase behavior of a linear PNiPAM with viscosity averaged molecular weight \mbox{$M_\mathrm{v}$ = 39 000 g/mol} at a concentration of $c$ = 10$^{-2}$ g/ml as a function of methanol molar fraction $X_\mathrm{MeOH}$. Lower panel: Dependence of the radius of gyration $R_\mathrm{g}$ of linear PNiPAM on $X_\mathrm{MeOH}$ adapted from data obtained by Zhang and Wu.\cite{Zhang2001}}
\end{figure}
 
In direct correlation to the observed re-entrance from a one-phase to a two-phase to a one-phase system PNiPAM exhibits a re-entrant coil-to-globule-to-coil transition as a function of solvent composition; this is sketched in the lower panel of Fig. 1, where we report the development of the PNiPAM dimensions as a function of $X_\mathrm{MeOH}$ at $T$ = $20\,^{\circ}{\rm C}$ adapted from the data obtained by Zhang and Wu.\cite{Zhang2001} In the range of solvent compositions, where we observe the appearance of precipitated phases, the PNiPAM chain exhibits a drastic conformational change from a fully swollen coil at low $X$ to a globular state at intermediate $X$ to again a fully swollen coil at high $X$.

Different approaches have been proposed to account for the coil-to-globule-to-coil transition and the re-entrant phase behavior. Although such behavior has been observed for several mixtures of water and organic solvents\cite{Costa2002,Hao2010,Hore2013}  most of the work is focused on the description of the phase behavior of \mbox{PNiPAM} in water/methanol mixtures.\cite{Schild1991,Zhang2001,Kojima2013,Tanaka2008,Winnik1992,Walter2012,Scherzinger2014,Mukherji2013,Pang2010} These approaches can be divided in two basic classes. In the first class, co-nonsolvency is considered to be the result of competitive adsorption.\cite{Kojima2013, Tanaka2008,Walter2012,Tanaka2009} Both water and methanol are thought to interact with PNiPAM by direct hydrogen bonds forming segments that are composed of either of the two solvents, these segments being separated by solvent free regions. At the solvent composition where the competition for such cooperative hydration is largest the solvent coverage of PNiPAM decreases drastically, resulting in the observed solubility gap. By contrast, in the second class of approaches the solvent structure itself is considered to be at the origin of co-nonsolvency. Zhang and Wu suggested that the co-nonsolvency behavior is due to the formation of water/methanol complexes that are poor solvents for PNiPAM.\cite{Zhang2001} Solubility of PNiPAM is then only given as long as there are sufficient free water or methanol molecules available for the hydration of PNiPAM; at stoichiometric compositions where all solvent molecules are involved in complexes PNiPAM is insoluble. Hao et al. studying the co-nonsolvency of PNiPAM in water/tetrahydrofuran mixtures again proposed that fluctuations in solvent composition would impact the solubility of PNiPAM; at the solvent composition where these fluctuations are largest PNiPAM would become insoluble.\cite{Hao2010}

The experimental evidence shown in this work denotes that the solvent state is indeed determining the phase behavior of PNiPAM in water-rich environments. However, instead of being controlled by the solvent structure we show that it is the energetic state of the solvent that is the determining parameter. We expose how such control by the solvent energetics implies that hydrophobic hydration is the predominant contribution controlling the phase behavior of PNiPAM in water-rich environments, the addition of alcohol to water gradually suppressing hydrophobic hydration. By contrast, in alcohol-rich environments the phase behavior of PNiPAM is governed by the classical polymer contributions to solution thermodynamics. The solvent-PNiPAM interactions are here largely nonspecific and the addition of water to alcohol worsens the solvent quality. Because adding alcohol to water suppresses hydrophobic hydration and adding water to alcohol worsens the solvent quality there is a range in solvent compositions where PNiPAM is essentially insoluble, which accounts for the co-nonsolvency behavior of PNiPAM in water/alcohol mixtures.

\section{\label{sec:setup}Experimental}
For our experiments we use both linear PNiPAM and PNiPAM microgels. The linear PNiPAM systems are purchased from Polymer Source Inc. and have differing viscosity averaged molecular weights $M_\mathrm{v}$ and polydispersities $M_\mathrm{w}$/$M_\mathrm{n}$, $M_\mathrm{w}$ and $M_\mathrm{n}$ denoting respectively the weight and number averaged molecular weights: PNiPAM 1 $M_\mathrm{v}$ = 39 000 g/mol and $M_\mathrm{w}$/$M_\mathrm{n}$ = 1.45; PNiPAM 2 $M_\mathrm{v}$ = 1 050 000 g/mol and $M_\mathrm{w}$/$M_\mathrm{n}$ = 1.4. The PNiPAM microgels are synthesized as described by Senff and Richtering.\cite{Senff1999} Due to the use of an ionic initiator the microgels are charged. In our studies probing the cloud point of the microgel systems these charges are screened by the addition of sodium thiocyanate, where we set the salt concentration to 0.03 M. In our studies probing the microgel dimensions we use solutions of PNiPAM microgels without added salt. The charged groups on the microgels provide sufficient electrostatic repulsion to prevent aggregation, which enables us to determine the microgel dimensions in the fully collapsed state without the need to use extremely low concentrations, as this is otherwise the case.\cite{Zhang2001,Wu1998} All our samples are prepared by mixing stock solutions of linear PNiPAM or PNiPAM microgels in respectively pure water (Milli-Q) and analytical grade alcohols, so to obtain the desired alcohol molar fraction $X$. As alcohols we use methanol, ethanol, isopropanol and propanol.

The critical solution temperatures $T_{c}$ of our PNiPAM samples are determined in cloud point measurements. Sealed glass tubes, containing the PNiPAM solutions, are placed in a homemade temperature cell, where the temperature can be controlled in a range of -$20\,^{\circ}{\rm C}$ to $60\,^{\circ}{\rm C}$ with a precision of ± $0.1\,^{\circ}{\rm C}$. After a first approximate assessment of $T_\mathrm{c}$ using a fast temperature ramp we approach the critical temperatures from below or above depending on whether the transition is characterized by a lower or upper critical solution temperature in steps of $0.1\,^{\circ}{\rm C}$; the solutions are allowed to equilibrate for at least 5 minutes at each temperature. The cloud point is determined by visually assessing the onset to turbidity. The critical solution temperatures obtained in control experiments using a commercial light scattering apparatus (ALV-5000) to measure the cloud point as the onset to a large scattering intensity are consistent with those determined visually.

The temperature-dependent dimensions of PNiPAM are characterized in static light scattering experiments. To facilitate our studies we mostly characterize the dimensions of the PNiPAM microgels; these have a significantly larger scattering cross section than linear PNiPAM systems, such that the angular dependent scattering intensity can be determined with a higher accuracy. The angular dependent scattering intensity, $I(q)$, is determined over a range of scattering wave vectors of $q$ = 8$\mu$m$^{-1}$ to $q$ = 30 $\mu$m$^{-1}$, and the radius of gyration $R_\mathrm{g}$ is determined by using the Guinier approximation \mbox{$I(q)$ = exp\Big\{-$ \frac{1}{3} $$q$$^{2}$$R_\mathrm{g}$$^{2}$\Big\}}.\cite{Guinier1955} Several studies revealed that the cross-linking density of PNiPAM microgels is inhomogeneous, the structure of the microgel being reasonably described as core-shell particles.\cite{Mason2005,Reufer2009,Stieger2004} The radius of gyration is here mainly a measure of the dimensions of the highly cross-linked core of the microgel, while the hydrodynamic radius, measured in dynamic light scattering experiments, is a measure of the overall dimensions of the microgels including the shell.\cite{Senff1999} We here consider the radius of gyration probing the almost evenly crosslinked core rather than the hydrodynamic radius to avoid possible artefacts due to the inhomogeneity of the microgels.

\section{\label{sec:results}Results and discussion}

Though the phenomenon of co-nonsolvency is rather rare in the general context of polymer solutions, for PNiPAM co-nonsolvency is ubiquitous and can be observed in various mixtures of water and organic solvents.\cite{Schild1991,Winnik1990,Costa2002,Hao2010} In these mixtures the re-entrance is characterized by either a lower critical solution temperature (LCST) or an upper critical solution temperature (UCST), while the transitions to globular and phase separated states observed in water-rich environments are always characterized by a LCST. As representative examples of these two classes of re-entrant behavior we here investigate the phase behavior of PNiPAM in water/methanol and water/ethanol mixtures in more detail. The re-entrance transition in water/methanol mixtures is characterized by a LCST,\cite{Schild1991,Winnik1990} while the re-entrance transition in water/ethanol mixtures is characterized by an UCST.\cite{Costa2002}

To understand whether these differences in the re-entrant behavior denote different solvation mechanisms for PNiPAM in respectively water/methanol and water/ethanol mixtures we investigate the critical solution temperatures $T_\mathrm{c}$ of different PNiPAM systems in both solvent mixtures as a function of alcohol molar fraction $X$. In particular, we explore the impact of PNiPAM concentration and molecular weight, as well as the effect of PNiPAM architecture on the $X$-dependence of $T_\mathrm{c}$, by investigating the phase behavior of both linear PNiPAM and PNiPAM microgels. Upon addition of small amounts of alcohol the LCST of the PNiPAM solutions initially decreases in both solvent mixtures, as shown in Fig. 2(a) and (b). However, the most striking feature of the phase behavior of PNiPAM at low $X$ is that $T_\mathrm{c}$ is essentially independent of concentration, molecular weight and architecture; this is evidenced by the almost perfect collapse of the four datasets, corresponding respectively to solutions of PNiPAM 1 (\mbox{$M_\mathrm{v}$ = 39 000 g/mol}) at concentrations of $c$ = 2$\cdot$10$^{-3}$ g/ml and \mbox{$c$ = 10$^{-2}$ g/ml}, PNiPAM 2 (\mbox{$M_\mathrm{v}$ = 1 050 000 g/mol}) at a concentration of \mbox{$c$ = 10$^{-2}$ g/ml} and PNiPAM microgels at a concentration of \mbox{$c$ = 8$\cdot$10$^{-4}$ g/ml}.

By contrast, in the re-entrant range of $X$, at larger $X$, the critical solution temperatures depend on all three parameters: PNiPAM concentration, molecular weight and architecture. This range of $X$ also defines the phase space where the development of the critical solution temperatures is qualitatively different for respectively water/methanol and water/ethanol mixtures. In water/methanol mixtures the LCST goes through a minimum and then increases again. In water/ethanol mixtures the LCST seemingly diverges to minus infinity at some critical solvent composition; beyond that composition there is a range of $X$ where PNiPAM is insoluble in the whole experimentally accessible temperature range of -20 to $60\,^{\circ}{\rm C}$; at even larger $X$ PNiPAM is then again soluble, this re-entrance being characterized by an UCST. Despite the differences in the type of critical solution temperature defining the re-entrant boundary, the re-entrance in both alcohol mixtures has a common characteristic, namely that the position of the boundary is a function of PNiPAM concentration and molecular weight, as well as PNiPAM architecture. Such emergence of a dependence of $T_\mathrm{c}$ on PNiPAM characteristics at larger $X$ suggests that the re-entrance has a common origin independent of whether we observe a LCST or UCST transition. 

\begin{figure}[H]
	\centering
	\includegraphics[scale=0.49]{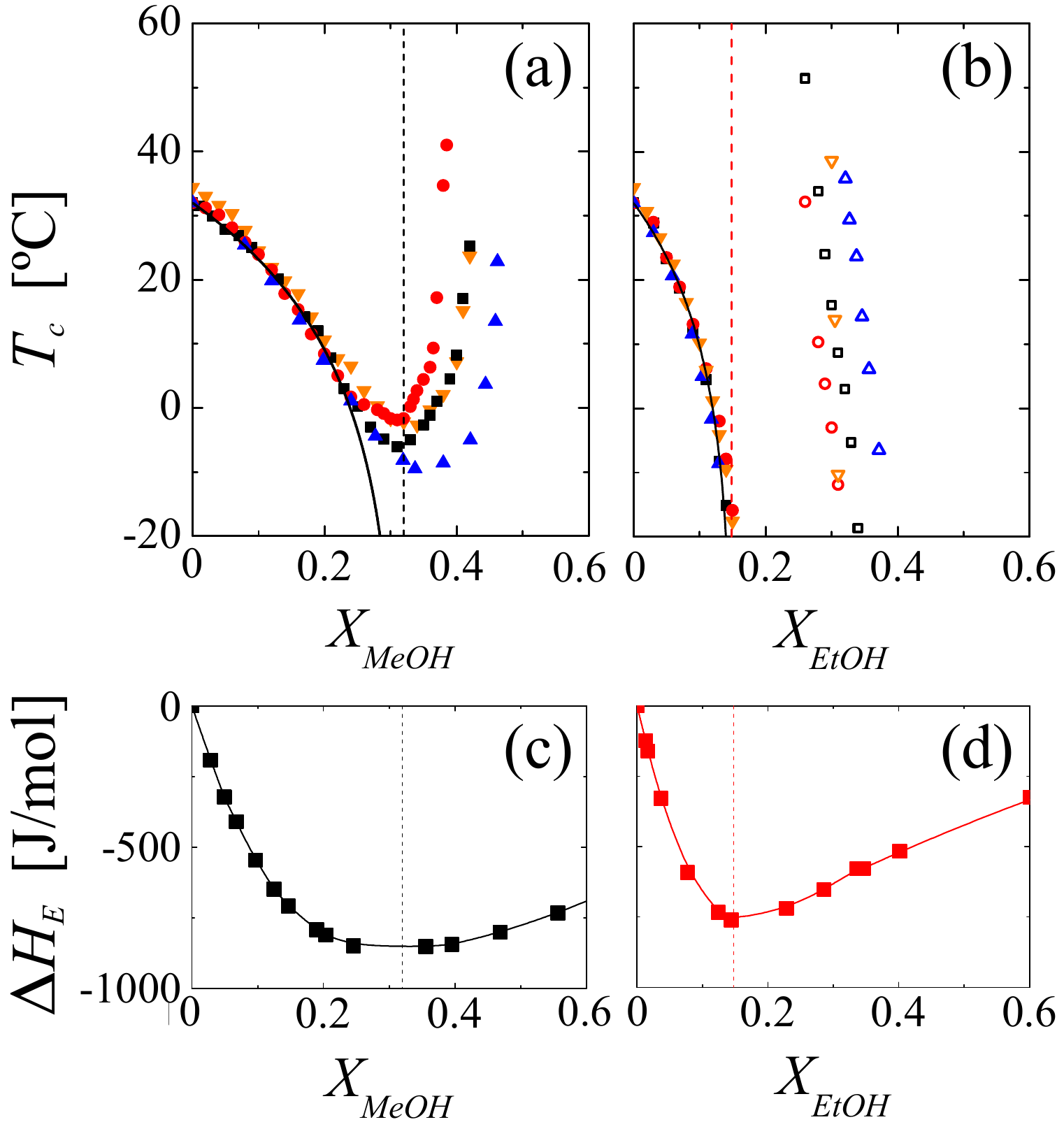}
	\caption{\label{Fig2}(a, b) Critical solution temperatures $T_\mathrm{c}$ of PNiPAM solutions in (a) water/methanol mixtures and (b) water/ethanol mixtures as a function of the alcohol molar fraction $X$. Full and open symbols denote respectively LCST and UCST. Black squares denote the critical solution temperatures of PNiPAM 1 ($M_\mathrm{v}$ = 39 000 g/mol) at \mbox{$c$ = 10$^{-2}$ g/ml}, red circles that of PNiPAM 1 at \mbox{$c$ = 2$\cdot$10$^{-3}$ g/ml}, blue triangles up that of PNiPAM 2 ($M_\mathrm{v}$ = 1 050 000 g/mol) at \mbox{$c$ = 10$^{-2}$ g/ml} and orange triangles down that of PNiPAM microgels at $c$ = 8$\cdot$10$^{4}$ g/ml. The solid lines correspond to approximations by  $T_\mathrm{c}$($X$) = $T_\mathrm{c}$($0$)\bigg(1-$\frac{X}{X*}$\bigg)$^{\alpha}$ with $\alpha$ = 0.07 and $T_\mathrm{c}$($0$) = 305 K for both solvent mixtures and with $X$*$_\mathrm{MeOH}$ = 0.32 and $X$*$_\mathrm{EtOH}$ = 0.15 for water/methanol and water/ethanol mixtures respectively. (c, d) Data obtained by Lama and Lu.\cite{Lama1965} Excess enthalpy of mixing, $\Delta$$H_\mathrm{E}$, of (c) water/methanol mixtures and (d) water/ethanol mixtures as measured at $T$ = $25\,^{\circ}{\rm C}$. The continuous lines are guides to the eye. The solvent compositions $X$* at which $\Delta$$H_\mathrm{E}$ exhibits a minimum are indicated by vertical dashed lines. Though the magnitude of $\Delta$$H_\mathrm{E}$ is a function of temperature, $X$* is independent of $T$.\cite{Bischofberger2014}}
\end{figure}

The differences in the dependence of $T_\mathrm{c}$ on PNiPAM characteristics at respectively low and high $X$ suggest a partitioning of phase space in two distinct regimes: a low $X$-regime where the classical mixing contributions to polymer solution thermodynamics are almost irrelevant for the phase behavior of PNiPAM, and a high $X$-regime where these contributions impact the phase behavior of PNiPAM. Indeed, in classical polymer solutions the phase behavior is strongly determined by the entropy of mixing $\Delta$$S_\mathrm{m}$, which depends on the volume fractions of the solvent and the polymer, $\phi$$_{1}$ and $\phi$$_{2}$, and on the constraints set by the polymer conformation, i.e. the degree of polymerization $N$ for linear polymers; \mbox{$\Delta$$S_\mathrm{m}$ = -$k_\mathrm{B}$[$\phi$$_{1}$ ln $\phi$$_{1}$ + ($\phi$$_{2}$/$N$) ln $\phi$$_{2}$]}, with $k_\mathrm{B}$ the Boltzmann constant.

The dependence of $T_\mathrm{c}$ on PNiPAM molecular weight and concentration observed at high $X$ in both water/methanol and water/ethanol mixtures is thus consistent with the behavior expected for classical polymer solutions; it indicates that the mixing entropy is here a significant contribution for the phase behavior of PNiPAM solutions. This applies to both type of transitions, the LCST transitions observed for PNiPAM solutions in water/methanol mixtures and the UCST transitions observed for PNiPAM solutions in water/ethanol mixtures. UCST transitions are those most commonly observed; they are due to the gain in mixing entropy that outplays the enthalpic contributions at high enough temperatures.\cite{Rubinstein2003} LCST transitions are rarer and somewhat less well understood within the context of classical polymer solutions; they can be assigned to a gain in free volume for the smaller solvent molecules upon phase separation that dominates over the loss in mixing entropy at higher temperatures.\cite{Patterson1969,Patterson1982,Somcynsky1982} This can be understood by considering that polymers have on themselves a smaller free volume than lower molecular mass compounds, such that the mixing of both effectively entails a loss of free volume for the lower molecular mass compound, i.e. the solvent molecules. In fact, the phase space of classical polymer solutions is generally considered to be characterized by both type of transitions, where LCST $>$ UCST.\cite{Patterson1969,Patterson1982,Somcynsky1982} The difference in the type of transition observed at larger $X$ indicates that our experimental temperature window covers the LCST-range of phase space for PNiPAM in water/methanol mixtures, while we probe the UCST-range for PNiPAM in water/ethanol mixtures. In both cases the phase behavior of PNiPAM qualitatively agrees with that expected for classical polymer solutions, which indicates that the solvent-polymer interactions are here essentially nonspecific, such that the concept of solvent quality applies.\cite{Rubinstein2003} The dependence of the critical solution temperatures with increasing $X$ can here be interpreted as that the solvent quality increases with increasing $X$.
 Indeed, both the increase of the LCST with increasing $X_\mathrm{MeOH}$ and the decrease of the UCST with increasing $X_\mathrm{EtOH}$ indicate that the phase space for homogeneous solutions widens with increasing $X$. In fact, the physics is here more properly captured by considering the behavior reversely: alcohols are good solvents for PNiPAM and the addition of water to alcohol worsens the solvent quality, as defined for classical solutions, where the solvent-polymer interactions are nonspecific.

In fact, the solubility of PNiPAM at low $X$ is not due to a good solvent quality in the classical sense, but to a fundamentally different mechanism driving the phase behavior of \mbox{PNiPAM} solutions in water-rich environments.\cite{Bischofberger2014} This mechanism is hydrophobic hydration, which can be understood as a pure solvent problem; the formation of a hydration shell around hydrophobic entities is enthalpically favorable for water at the cost of being entropically unfavorable.\cite{Ball2008,Southall2002,Lamanna1996} Upon increasing the temperature the entropic gain obtained by releasing the water molecules from the hydration shell drives phase separation, which naturally leads to a LCST transition. First evidence that hydrophobic hydration is the prevailing contribution governing the LCST behavior of PNiPAM at low $X$ is the independence of the LCST on PNiPAM concentration, molecular weight and architecture. Indeed, considering the contributions of hydrophobic hydration to solution thermodynamics as being solely set by whether or not the presence of \mbox{PNiPAM} is favorable to the energetic state of water,\cite{Ball2008,Southall2002,Lee1996,Moelbert2003,Muller1990} we do not expect the phase transition temperatures to depend on the polymer contributions that usually govern the phase behavior of classical polymer solutions, consistent with the observed behavior.

The impact of adding small amounts of alcohol to aqueous solutions of PNiPAM can be understood within the concept of the kosmotropic effect.\cite{Bischofberger2014} Indeed, alcohols are known to be kosmotropic agents, agents that are presumed to strengthen the hydrogen-bonded network of water without disrupting it.\cite{Galinski1997,Moelbert2004} As the water enthalpy decreases upon addition of alcohol, the gain for water to form a hydration shell around PNiPAM decreases. This leads to a decrease of the LCST with increasing $X$, as the gain in water entropy upon release of the water molecules from the hydration shell dominates at lower temperature. That this scenario accounts for the phase behavior of PNiPAM in the low $X$-regime can be inferred from direct correlations between the excess enthalpy of mixing of the water/alcohol mixtures and the development of the LCST of PNiPAM in these mixtures. As shown in Fig. 2(c) and (d), the excess enthalpy of mixing $\Delta$$H_\mathrm{E}$ of the water/alcohol mixtures exhibits a minimum at a given solvent composition $X$*.\cite{Lama1965} This composition is larger for the water/methanol mixtures than for the water/ethanol mixtures, reminiscent of the development of the LCST of the PNiPAM solutions with $X$: the LCST decreases more slowly with $X$ for PNiPAM in water/methanol mixtures than for PNiPAM in water/ethanol mixtures. In fact, using $X$* to approximate the initial decrease of the LCST with a critical-like function of the form \mbox{$T_\mathrm{c}$($X$) = $T_\mathrm{c}$($0$)\bigg(1-$\frac{X}{X*}$\bigg)$^{\alpha}$} yields a reasonable description of the data in the low $X$-regime, as shown in Fig. 2(a) and (b). Such correlation between thermodynamic characteristics of the solvent mixture and the PNiPAM solution in these mixtures is found for PNiPAM solutions in other aqueous mixtures containing organic solutes that belong to the class of kosmotropes.\cite{Bischofberger2014} This denotes the energetic state of water as key parameter controlling the phase behavior of PNiPAM at low $X$, consistent with the notion that hydrophobic hydration is the determining contribution in the low $X$-regime. Indeed, our findings can be considered as an experimental proof of the validity of the concepts used in two-state models, where solely the difference in energy between bulk water and the water forming a hydration shell around a hydrophobic entity (shell water) is considered relevant for the description of hydrophobic hydration.\cite{Lee1996,Moelbert2003,Muller1990}

Within this framework the solvent composition $X$* at which $\Delta$$H_\mathrm{E}$ becomes minimal denotes the solvent composition at which the presence of kosmotropes fully optimizes the energetic state of bulk water, such that the difference between bulk and shell water becomes zero. Indeed, previous experiments probing the enthalpy change $\Delta$$H_\mathrm{E}$ associated with the LCST transition of PNiPAM solutions in the low $X$-regime revealed that the decrease of the LCST with increasing $X$ correlates with a decrease of $\Delta$$H_\mathrm{E}$,\cite{Schild1991,Bischofberger2014} the extrapolation of $\Delta$$H_\mathrm{E}$ to zero effectively denoting $X$* as the limit to hydrophobic hydration.\cite{Bischofberger2014}

Thus, while in the high $X$-regime an increase in the water content leads to a decrease in the solubility of PNiPAM due to a decrease in the solvent quality, in the low $X$-regime an increase in alcohol content leads to a decrease in the gain for water to form a hydration shell around the hydrophobic groups of PNiPAM due to the kosmotropic effect. Because of these opposing trends PNiPAM is insoluble at intermediate $X$, which accounts for the observed co-nonsolvency behavior. However, the composition at which we observe the transition between the hydrophobic hydration determined regime and the regime where the classical mixing contributions prevail is not uniquely defined. This transition will sensitively depend on the range of $X$, where a given PNiPAM system still forms a homogeneous solution due to the classical mixing contributions, which in turn depends on the PNiPAM concentration, molecular weight\cite{Tanaka2011} and architecture.

To explore how the distinct solvation mechanisms impact the coil-to-globule-to-coil transition or respectively the volume phase transition of PNiPAM microgels we investigate the dimensions of the PNiPAM microgels as a function of $X$ in more detail. Our choice of system is here motivated by the fact that PNiPAM microgels are colloidally stable in the collapsed phase, due to charged groups that are introduced by the charged initiator used in the synthesis of these systems (see experimental section). Additional experiments probing the dimensions of linear PNiPAM below the LCST or respectively above the UCST are shown in the supporting information. The trend observed in the re-entrant volume transitions of microgels is broadly consistent with the coil-to-globule-to-coil transition observed for linear PNiPAM.\cite{Zhang2001} At a fixed temperature of $T$ = $12.5\,^{\circ}{\rm C}$ an increase in the alcohol molar fraction initially induces a decrease in the radius of gyration $R_\mathrm{g}$ of the microgels, as shown in Fig. 3(a). This decrease is then followed by an increase, where $R_\mathrm{g}$ eventually reaches a final value that is slightly below that of the microgel in pure water. 

\begin{figure}[H]
	\centering
	\includegraphics[scale=0.46]{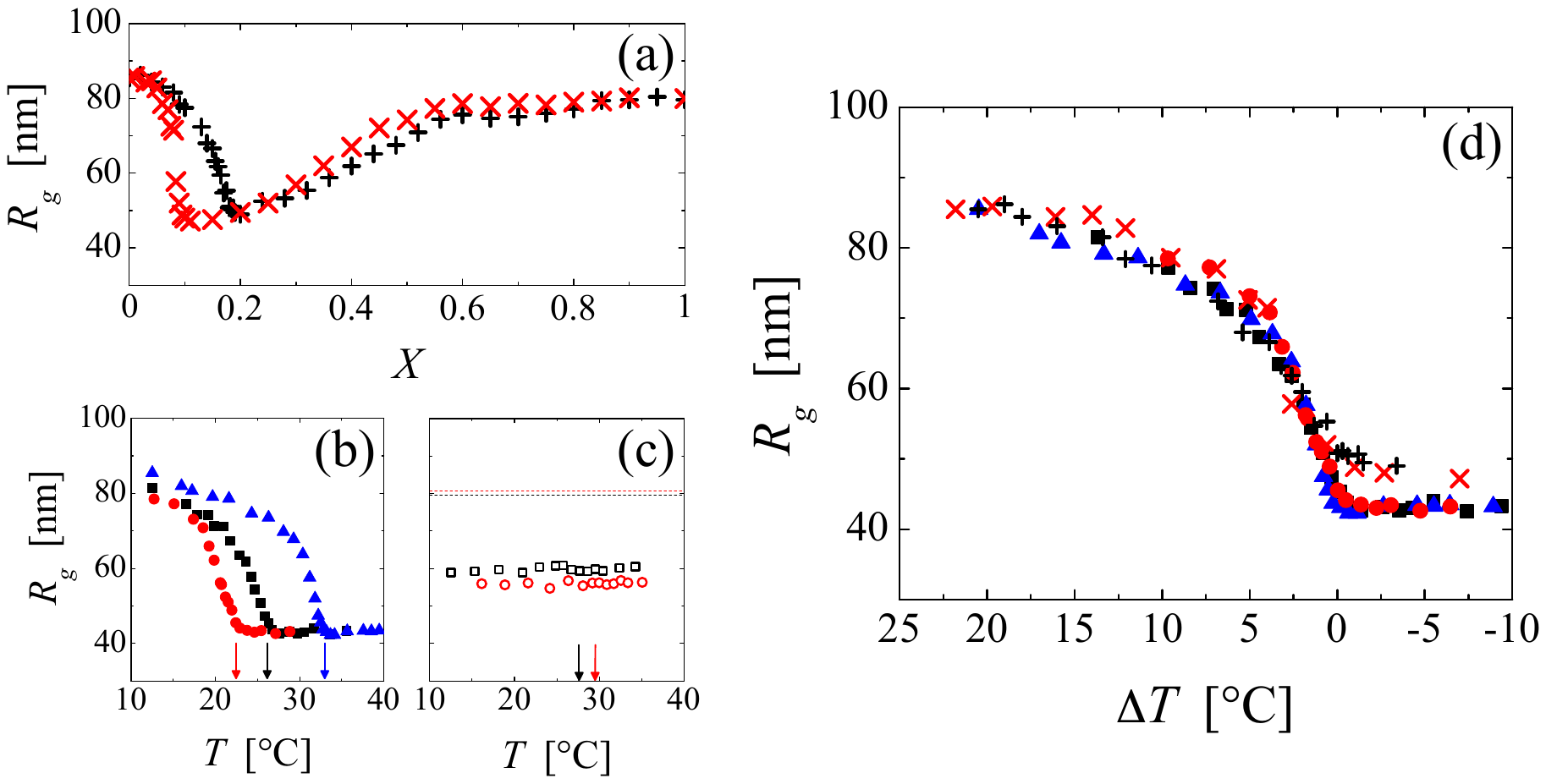}
	\caption{\label{Fig3} Radius of gyration $R_\mathrm{g}$ of PNiPAM microgels at a concentration of 
$c$ = 2.4$\cdot$10$^{-6}$ g/ml. (a) Dependence of $R_\mathrm{g}$ on the alcohol molar fraction $X$ at a fixed temperature of $T$ = $12.5\,^{\circ}{\rm C}$ for water/methanol mixtures (black pluses) and water/ethanol mixtures (red crosses). (b) Temperature dependence of $R_\mathrm{g}$ in the low $X$-regime: $X$ = 0 (blue triangles), $X_\mathrm{MeOH}$ = 0.08 (black squares) and $X_\mathrm{EtOH}$ = 0.06 (red circles). PNiPAM exhibits a coil-to-globule transition that coincides with the LCST, which is denoted by arrows. (c) Temperature dependence of $R_\mathrm{g}$ in the high $X$-regime: $X_\mathrm{MeOH}$ = 0.37 (black squares) and $X_\mathrm{EtOH}$ = 0.275 (red circles). The arrows indicate the critical solution temperatures, corresponding to a LCST for $X_\mathrm{MeOH}$ = 0.37 and to an UCST for $X_\mathrm{EtOH}$ = 0.275. Dotted black and red lines denote respectively the magnitudes of the microgel radii in pure methanol and ethanol. (d) Reporting $R_\mathrm{g}$ as a function of the reduced temperature $\Delta$$T$ = $T_\mathrm{c}$-$T$ collapses all data obtained for $X$ $<$ $X$*. As in (a) the black pluses and red crosses denote the data obtained at a fixed temperature of $T$ = $12.5\,^{\circ}{\rm C}$ and varying $X$. As in (b) the full symbols denote the data obtained at fixed $X$ and varying temperatures.
}
\end{figure}

To assess how the set temperature affects the observed behavior we determine the temperature dependence of the microgel dimensions at some fixed solvent compositions, selecting systems for which the critical solution temperature is conveniently accessed in experiments. As examples of systems with \mbox{$X$ $<$ $X$*} we investigate $X$ = 0 ($T_\mathrm{c}$ = $33.0\,^{\circ}{\rm C}$), \mbox{$X_\mathrm{MeOH}$ = 0.08} ($T_\mathrm{c}$ = 26.2$\,^{\circ}{\rm C}$), $X_\mathrm{EtOH}$ = 0.06 ($T_\mathrm{c}$ = $22.4\,^{\circ}{\rm C}$); representative of systems with $X$ $>$ $X$* we investigate \mbox{$X_\mathrm{MeOH}$ = 0.37} (\mbox{$T_\mathrm{c}$ = 27.5$\,^{\circ}{\rm C}$}) and $X_\mathrm{EtOH}$ = 0.275 ($T_\mathrm{c}$ = $29.5\,^{\circ}{\rm C}$). Remarkably, while the dimensions of the systems with \mbox{$X$ $<$ $X$*} exhibit a strong temperature dependence upon approach of $T_\mathrm{c}$, the dimensions of the systems with $X$ $>$ $X$* are temperature independent across the phase separation boundary, as shown in respectively Fig. 3(b) and (c); the critical solution temperatures characterizing the boundary to phase separated states are here indicated by arrows.

In the low $X$-regime the functional development of $R_\mathrm{g}$ with $T$ is similar to that in pure water. Far enough from the LCST the PNiPAM dimensions are almost independent of $T$. Upon increasing the temperature $R_\mathrm{g}$ decreases, exhibiting dimensions that are governed by the relative distance of the temperature to the LCST, $\Delta$$T$ = $T_\mathrm{c}$-$T$. This becomes most evident when reporting $R_\mathrm{g}$ as a function of $\Delta$$T$. As shown in Fig. 3(d), all data obtained for $X$ $<$ $X$* including those shown in Fig. 3(a) and (b) collapse to a unique master curve. Deviations from the master curve are only observed for solvent compositions very near to $X$*. This collapse shows that in the low $X$-regime the development of the dimensions with $X$ or $T$ are uniquely set by $\Delta$$T$. Within the context of hydrophobic hydration our findings effectively denote that the hydration shell of PNiPAM remains unaffected by the addition of small amounts of alcohols; the dimensions of PNiPAM are essentially independent of $X$ for $T$ $<<$ LCST. For a fixed $T$ solely the shift in the LCST leads to a decrease in the dimension with increasing $X$.

It is worth emphasizing that the coil-to-globule transition temperatures of linear PNiPAM or respectively the volume phase transition temperature of PNiPAM microgels, denoting the transition from a swollen state at low temperature to a collapsed state at high temperature, always coincide with the phase separation transition temperatures $T_\mathrm{c}$ in the low $X$-regime. Such coincidence is not generally expected. In classical polymer solutions $T_\mathrm{c}$ is a function of the polymer concentration, while the coil-to-globule transition temperature $T_\mathrm{c-g}$ is fixed, $T_\mathrm{c-g}$ and $T_\mathrm{c-g}$ therefore do not normally coincide.\cite{Rubinstein2003,Swislow1980} The coincidence observed here effectively reflects the same physics as the independence of $T_\mathrm{c}$ on PNiPAM concentration, molecular weight and architecture; namely that the entropic contributions of the polymer are basically irrelevant for the phase behavior of PNiPAM in water-rich environments, the phase behavior being instead governed by the enthalpic gain for water to hydrate the hydrophobic groups of PNiPAM.

In the high $X$-regime, where we consider the solvent-PNiPAM interactions to be nonspecific, it is tempting to assign the increase in the PNiPAM dimensions with increasing $X$ to the increase in solvent quality inferred from the $X$-dependence of the LCST and UCST. However, the complete independence of the PNiPAM dimensions on temperature across the phase separation boundary is somewhat surprising. As shown in the supporting information, this independence is robust and can be found for linear PNiPAM as well. Indeed, though we do not expect the coil-to-globule transition temperatures to directly coincide with the phase transition temperatures in the high $X$-regime, our temperature scans around $T_\mathrm{c}$ are quite wide and our findings seemingly indicate that there is no coil-to-globule transition at all within the range of $T_\mathrm{c}$. This is consistent with recent experiments probing the conformation of linear PNiPAM in time-resolved anisotropy measurement; there the segmental mobility of the chain was found to exhibit a strong decrease at $T_\mathrm{c}$ for PNiPAM in the low $X$-regime, while it remained constant across the phase separation boundary for PNiPAM in the high $X$-regime. This indicates that PNiPAM does not exhibit any coil-to-globule transition near $T_\mathrm{c}$ in the high $X$-regime.\cite{Chee2011} To date we do not have a conclusive explanation for this behavior. Nonetheless, let us here emphasize that the emergence of temperature-independent dimensions in the re-entrant range of $X$ is independent of whether the re-entrance is characterized by a LCST or an UCST. This corroborates that the origin of the re-entrance is common in solvent mixtures exhibiting either type of reentrance and precludes models accounting for the co-nonsolvency of PNiPAM exclusively for re-entrant transitions that are characterized by a LCST.\cite{Zhang2001,Kojima2013,Tanaka2008,Tanaka2009}

Finally, it is worth stressing that the re-entrant globule-to-coil transition observed in the high $X$-regime is of very different nature than the coil-to-globule transition observed at low $X$. While in the low $X$-regime $R_\mathrm{g}$ is set by $\Delta$$T$, in the high $X$-regime $R_\mathrm{g}$ is set by $X$. The development of the PNiPAM dimensions reported in Fig. 3(a) is thus unique in the high $X$-regime for a given PNiPAM system, while it depends on the set temperature in the low $X$-regime. This entails that the range of $X$ over which PNiPAM exhibits a globular state varies with temperature, which excludes scenarios accounting for the coil-to-globule-to-coil transition as being due to the formation of water/alcohol complexes with a precise stoichiometry.\cite{Zhang2001}

To further progress in assessing the parameters governing the phase behavior of PNiPAM in respectively the low and high $X$-regime we take advantage of our investigations probing the critical solution temperatures and PNiPAM dimensions in different solvent mixtures. In Fig. 4 we report the dependences of $T_\mathrm{c}$ and $R_\mathrm{g}$ on solvent composition including data obtained for PNiPAM in water/isopropanol and water/propanol mixtures. In these solvent systems the co-nonsolvency behavior is qualitatively similar to that of \mbox{PNiPAM} in water/ethanol; the re-entrance is characterized by an UCST behavior. Remarkably, reporting $T_\mathrm{c}$ and $R_\mathrm{g}$ as a function of $X$/$X$* leads to an almost prefect collapse of all data sets in the low $X$-regime, while the off-shift between the different data sets in the high $X$-regime becomes more significant, as shown in respectively Figure 4(b) and (e). By contrast, reporting the data as a function of the alcohol volume fraction $\phi$$_{a}$ leads to a reasonable collapse of all data in the high $X$-regime, while off-shifting the data in the low $X$-regime, as shown in respectively Fig. 4(c) and (f).

\begin{figure}[htb]
	\centering
	\includegraphics[scale=0.52]{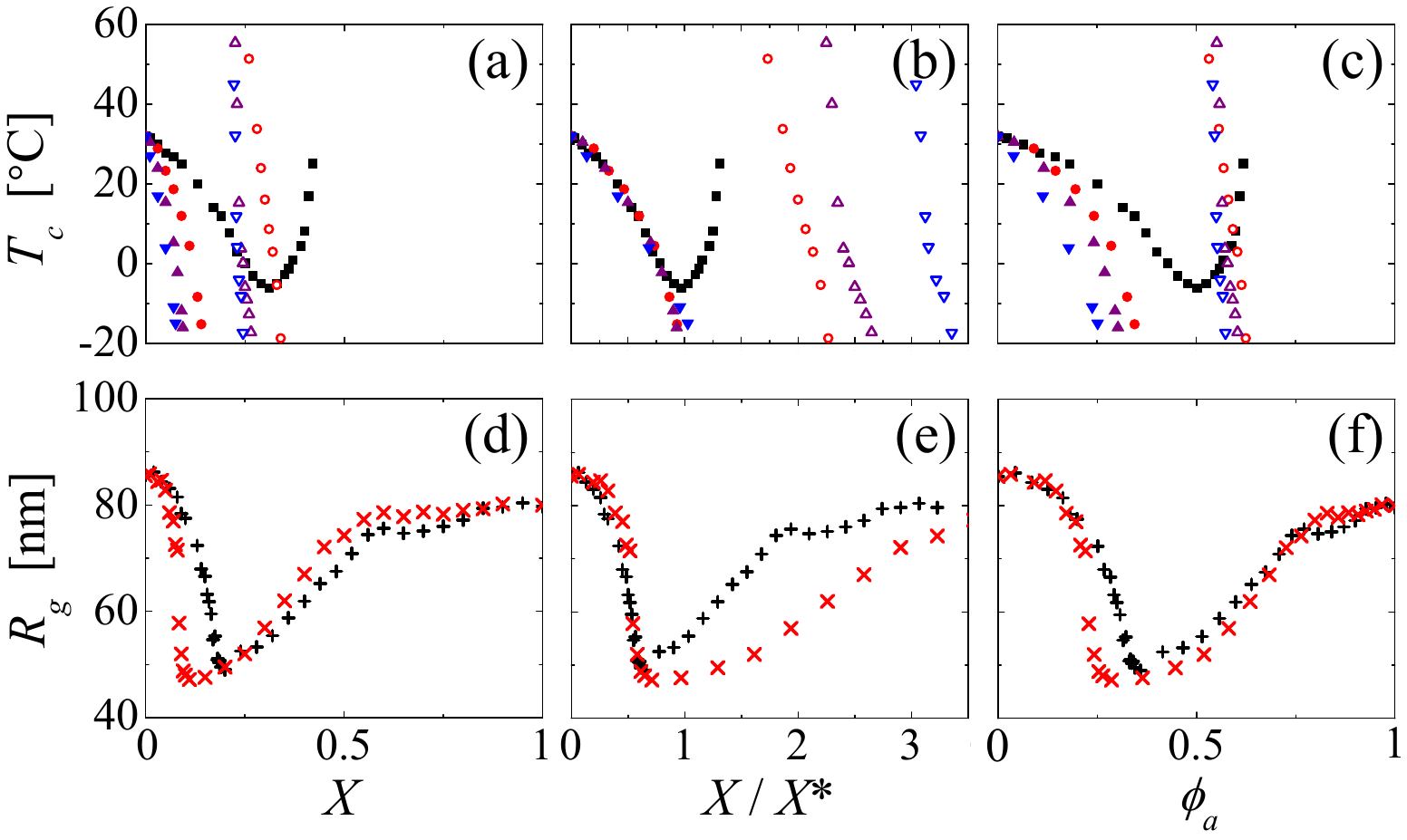}
	\caption{\label{Fig4}(a-c) Dependence of the critical solution temperature $T_\mathrm{c}$ on solvent composition for a fixed concentration and molecular weight of a linear PNiPAM, $c$ = 10$^{-2}$ g/ml and \mbox{$M_\mathrm{v}$ = 39 000 g/mol} \mbox{(PNiPAM 1)}. The solvent mixtures are water/methanol (black squares), water/ethanol (red circles), water/isopropanol (purple triangles up) and water/propanol mixtures (blue triangles down). Full and open symbols denote respectively LCST and UCST. (d-f) Dependence of PNiPAM microgel dimensions on solvent composition at a fixed temperature of $T$ = $12.5\,^{\circ}{\rm C}$ in water/methanol (black pluses) and water/ethanol mixtures (red crosses). For both series of experiments reporting the data as a function of $X$/$X$* leads to a collapse of the different data sets in the low $X$-regime, while reporting the data as a function of $\phi$$_{a}$ leads to a collapse of the different data sets in the high $X$-regime.}
\end{figure}

To understand the scaling with $X$* in the low $X$-regime let us recall that $X$* defines the solvent composition at which the excess mixing enthalpy of the water/alcohol mixtures exhibits a minimum independent of temperature: $X$*$_\mathrm{MeOH}$ = 0.32, $X$*$_\mathrm{EtOH}$ = 0.15, $X$*$_\mathrm{IsoPrOH}$ = 0.10 and $X$*$_\mathrm{PrOH}$ = 0.07.\cite{Bischofberger2014,Lama1965} The scaling of the LCST with $X$/$X$* obtained for all water/alcohol systems under investigation clearly shows that the phase behavior of PNiPAM is controlled by the energetic state of the solvent. Considering that in the low $X$-regime alcohols solely impact the water state, $X$* can be taken as a gauge for the efficiency of a given alcohol to decrease the bulk water enthalpy; the LCST evolves relative to the composition $X$* at which the water enthalpy becomes minimal. The scaling of the LCST with $X$/$X$* directly infers that $R_\mathrm{g}$ scales with $X$/$X$*; because in the low $X$-regime the PNiPAM dimensions are a function of $\Delta$$T$ = $T_\mathrm{c}$-$T$ and $T_\mathrm{c}$ is a function of $X$/$X$*, the PNiPAM dimension obtained at a fixed $T$ scales with $X$/$X$*, consistent with the scaling behavior observed in Fig. 4(e).

The scaling of the critical solution temperatures with the alcohol volume fraction $\phi$$_{a}$ in the high $X$-regime can be understood within the Flory-Huggins theory of ternary mixtures,\cite{Rubinstein,Altena1982,Flory1953} if we assume that the solvent quality does not vary much from one alcohol to another. Let us note that all data reported in Fig. 4(a-c) are obtained for a given PNiPAM system at a fixed concentration, PNiPAM 1 (\mbox{$M_\mathrm{v}$ = 39 000 g/mol)} and $c$ = 10$^{-2}$ g/ml. Thus, the polymer volume fraction and molecular weight are fixed. Under the assumption that the geometry and size of the alcohols do not significantly alter the Flory-Huggins parameter that accounts for the PNiPAM-alcohol interactions the volume fraction of alcohol is then the only variable in the experiment, consistent with the observed scaling behavior. To evaluate whether such assumption is reasonable we consider the surface tensions between air and the different alcohols as a measure of the interactions between PNiPAM and the alcohols. Indeed, for the alcohols considered here the surface tensions are almost identical.\cite{Vazquez1995} Further supporting that the PNiPAM-alcohol interactions do not strongly depend on the alcohol used we find that the microgel dimensions are identical in pure methanol, ethanol, isopropanol and propanol. Thus, the scaling with $\phi$$_{a}$ obtained for a given PNiPAM molecular weight and concentration effectively corroborates that the solvent-PNIPAM interactions are largely nonspecific in the high $X$-regime, such that the phase behavior of PNiPAM at high $X$ can be explained within the frame of classical polymer solution models, where the volume fractions of respectively water, alcohol and \mbox{PNiPAM} are relevant parameters.\cite{Altena1982,Flory1953}

\section{\label{sec:results}Conclusions}
Our investigations probing the phase behavior of PNiPAM in different water/alcohol mixtures show that the phenomenon of co-nonsolvency and the related coil-to-globule-to-coil transition observed for PNiPAM in water/alcohol mixtures are due to two distinct mechanisms governing the phase behavior of PNiPAM in respectively water-rich and alcohol-rich environments.

The phase behavior of PNiPAM in water-rich environments is predominantly controlled by hydrophobic hydration, which in turn is governed by the enthalpy difference between bulk water and the water forming a hydration shell around the hydrophobic groups of PNiPAM.\cite{Bischofberger2014} Adding alcohol to water decreases the enthalpy of the bulk water due to the kosmotropic effect. This leads to a decrease in the enthalpy difference between bulk and shell water, which eventually vanishes at the solvent composition where the bulk water enthalpy is minimal.\cite{Bischofberger2014} This condition sets a well-defined limit to hydrophobic hydration.
 
In the alcohol-rich regime the phase behavior of PNiPAM is set by the classical mixing contributions to the thermodynamics of polymer solutions,\cite{Rubinstein2003} independent of whether the boundary to phase separation is characterized by a lower or upper critical solution temperature. The solvent-PNiPAM interactions are here to be considered as nonspecific; alcohols are good solvents for PNiPAM and in the context of nonspecific interactions the addition of water to alcohol worsens the solvent quality. This eventually leads to phase separation when a certain water content is exceeded.
 
Because an increase in alcohol content both suppresses hydrophobic hydration and increases the solvent quality, \mbox{PNiPAM} exhibits the phenomenon of co-nonsolvency in water/alcohol mixtures. Let us note that this scenario implies that co-nonsolvency should be regarded as a coarse grained phenomenon, where the detailed configuration of the solvent around the polymer\cite{Mukherji2013,Pang2010,Deshmukh2012} is less important than the mean energetics of the solvent and the solution in respectively the low and high $X$-regime. Since the effects of water-mediated solvation and mixing-determined solvation are not specific to PNiPAM in water/alcohol mixtures, we expect that this scenario also accounts for the co-nonsolvency of other amphiphilic polymers in binary solvent mixtures where one of the components is water. We hope that the conceptual framework proposed in this paper will stimulate the development of theoretical descriptions that combine the concepts used to describe hydrophobic hydration in terms of a two-state problem\cite{Lee1996,Moelbert2003,Muller1990} with those used to describe polymer solutions in binary fluid mixtures\cite{Altena1982,Flory1953} to quantitatively account for the co-nonsolvency phenomenon. 

\balance

\section*{Acknowledgements}
We thank Paolo De Los Rios and Tom Witten for illuminating discussions. Financial support from the Swiss National Science Foundation (grant numbers 200020-130056 and 200020-140908) and the Adolphe Merkle Foundation are gratefully acknowledged.\\

Electronic Supplementary Information available: Temperature dependence of radius of gyration of linear PNiPAM in water/alcohol mixtures.

\footnotesize{
\bibliography{cononsolvency}
}

\end{document}